# *Non-orthogonal multiple-relaxation-time lattice Boltzmann method for incompressible thermal flows*

Qing Liu[a], Ya-Ling He[a], Dong Li[a], Qing Li[b]

[a]*Key Laboratory of Thermo-Fluid Science and Engineering of Ministry of Education, School of Energy and Power Engineering, Xi'an Jiaotong University, Xi'an, Shaanxi, 710049, China*

[b]*School of Energy Science and Engineering, Central South University, Changsha 410083, China*

**Abstract**

In this paper, a non-orthogonal multiple-relaxation-time (MRT) lattice Boltzmann (LB) method for simulating incompressible thermal flows is presented. In the method, the incompressible Navier-Stokes equations and temperature equation are solved separately by two different MRT-LB equations, which are developed based on non-orthogonal basis vectors obtained from the combinations of the lattice velocity components. The macroscopic governing equations of incompressible thermal flows can be recovered from the method through the Chapman-Enskog analysis in the incompressible limit. Numerical simulations of several typical two-dimensional problems are carried out to validate the proposed method. It is found that the present results are in good agreement with the analytical solutions and/or other numerical results reported in the literature. Furthermore, the non-orthogonal MRT-LB model shows better numerical stability in comparison with the BGK-LB model, and the grid convergence tests indicate that the present MRT-LB method has a second-order convergence rate in space.

1. Introduction

The lattice Boltzmann (LB) method [1-3], as a mesoscopic numerical approach based on the kinetic theory, has achieved great success in simulating fluid flow and heat transfer problems in the last two decades [4-11]. Historically, the LB method originated from the lattice-gas automata (LGA) method [12], which can be viewed as a simplified fictitious molecular dynamics (MD) method utilizing discrete lattice, discrete time, and discrete particle velocities. Later He and Luo [13] demonstrated that the LB equation can be rigorously obtained from the continuum Boltzmann equation for single-particle distribution function by using a small Mach number expansion. Unlike the traditional numerical methods based on a direct discretization of the macroscopic continuum equations, the LB method is based on the mesoscopic kinetic equation for single-particle distribution function. Owing to its kinetic background, the LB method has some attractive advantages over the traditional numerical methods [4, 14]: (i) non-linearity (the collision process) is local and non-locality (the streaming process) is linear, whereas the transport term $\mathbf{u}\cdot\nabla\mathbf{u}$ in the Navier-Stokes (N-S) equations is non-linear and non-local at a time; (ii) the pressure of the LB method is simply calculated by an equation of state, while in traditional numerical methods it is usually necessary and costly to solve a Poisson equation for the pressure field of the incompressible N-S equations; (iii) complex boundary conditions can be easily formulated in terms of the elementary mechanical rules; (iv) nearly ideal for parallel computing with very low communication/computation ratio.

In the LB community, the Bhatnagar-Gross-Krook (BGK) LB model [15, 16] is still the most frequently used model because of its extreme simplicity. Since its introduction 25 years ago, the BGK-LB method has been widely employed to study incompressible isothermal and thermal flows. However, the simplicity of the BGK-LB model comes at the expense of inaccuracy in implementing boundary conditions [17, 18] and numerical instability at low viscosity [19]. The above mentioned shortcomings of the BGK-LB model can be easily addressed by using the multiple-relaxation-time (MRT) model proposed by d'Humières [20] in 1992, which is an important extension of the relaxation LB method developed by Higuera et al. [2, 3]. In the MRT model, the collision process is mapped onto the moment space through a transformation matrix, while the streaming process is still executed in the velocity space. It has been widely accepted that the MRT collision model can significantly improve the numerical stability of the LB schemes by carefully separating the relaxation rates of the hydrodynamic (conserved) and kinetic (non-conserved) moments [19, 21-24]. In recent years, the MRT-LB method has also been used to simulate convective heat transfer problems in the framework of the double-distribution-function (DDF) approach [25-29]. The DDF-based MRT-LB method [25-29] utilizes two different MRT-LB equations: a D2Q9 MRT-LB equation for the flow field and a D2Q5 MRT-LB equation for the temperature field. The DDF-based MRT-LB method for incompressible thermal flows with the Boussinesq approximation has a second-order rate of convergence in space [27], and it is expected to exhibit better numerical stability at low thermal diffusivity than the BGK-LB method.

To the authors' knowledge, the existing DDF-based MRT-LB models for incompressible thermal flows are developed based on orthogonal basis vectors obtained from the combinations of the lattice velocity components, i.e., the transformation matrix is an orthogonal one which can be constructed via the Gram-Schmidt orthogonalization process. From this point of view, most of the existing MRT-LB models can be viewed as classical orthogonal MRT-LB schemes. It is noted that the transformation matrix of the MRT-LB model is not necessary to be an orthogonal one [30-32]. The MRT-LB model can be developed based on non-orthogonal transformation matrix (non-orthogonal MRT-LB model), and therefore the Gram-Schmidt orthogonalization process is not needed. As pointed out by Geier et al.

[32], the non-orthogonal MRT-LB model for incompressible isothermal flows provides solutions that are asymptotically consistent with the N-S equations to second-order in diffusive scaling. Compared with the orthogonal transformation matrix, the non-orthogonal transformation matrix contains more zero elements, and it is expected that the non-orthogonal MRT-LB model is computationally more efficient than the orthogonal MRT-LB model. Hence, the aim of this work is to present a non-orthogonal MRT-LB method for simulating incompressible thermal flows in the framework of the DDF approach. Numerical simulations of several typical two-dimensional (2D) problems are carried out to validate the present method. Moreover, the spatial accuracy and numerical stability of the present method are also discussed.

The rest of this paper is organized as follows. In Section 2, the 2D non-orthogonal MRT-LB method for incompressible thermal flows is presented in detail. In Section 3, numerical simulations of several typical 2D problems, including the fully developed channel flow driven by a constant body force, the lid-driven cavity flow, the thermal flow in a channel with wall injection, and the natural convection flow in a square cavity, are carried out to validate the present MRT-LB method. Finally, some conclusions are made in Section 4.

## 2. Non-orthogonal MRT-LB method

In this section, a non-orthogonal MRT-LB method for incompressible thermal flows is presented in detail. In the method, two MRT-LB models for the incompressible N-S equations and temperature equation are developed based on non-orthogonal basis vectors. For incompressible thermal flows, the macroscopic governing equations can be expressed as follows:

$$\nabla \cdot \mathbf{u} = 0, \tag{1}$$

$$\frac{\partial \mathbf{u}}{\partial t} + \mathbf{u} \cdot \nabla \mathbf{u} = -\frac{1}{\rho_0}\nabla p + \upsilon \nabla^2 \mathbf{u} + \mathbf{F}, \tag{2}$$

$$\frac{\partial T}{\partial t} + \mathbf{u}\cdot\nabla T = \nabla\cdot(\alpha\nabla T), \tag{3}$$

where $\mathbf{u}$, $p$ and $T$ are the velocity, pressure and temperature of the fluid, respectively, $\rho_0$ is the reference density, $\upsilon$ is the kinematic viscosity, $\alpha$ is the thermal diffusivity, and $\mathbf{F} = (F_x, F_y)$ is the body force due to the presence of external force fields. Based on the Boussinesq approximation, the body force $\mathbf{F}$ is given by

$$\mathbf{F} = g\beta(T-T_0)\mathbf{j} + \mathbf{a}, \tag{4}$$

where $g$ is the gravitational acceleration, $\beta$ is the thermal expansion coefficient, $T_0$ is the reference temperature, $\mathbf{j}$ is the unit vector in the *y*-direction, and $\mathbf{a}$ is the body force induced by an external force.

*2.1 MRT-LB model for the flow field*

In this subsection, a D2Q9 MRT-LB model for the flow field is presented. The nine discrete velocities $\{\mathbf{e}_i | i = 0, 1, \ldots, 8\}$ of the D2Q9 lattice are given by [16]

$$\mathbf{e}_i = \begin{cases} (0,0), & i = 0, \\ \left(\cos[(i-1)\pi/2], \sin[(i-1)\pi/2]\right)c, & i = 1 \sim 4, \\ \left(\cos[(2i-9)\pi/4], \sin[(2i-9)\pi/4]\right)\sqrt{2}c, & i = 5 \sim 8, \end{cases} \tag{5}$$

where $c = \delta_x/\delta_t$ is the lattice speed, in which $\delta_t$ and $\delta_x$ are the discrete time step and discrete lattice spacing, respectively. The lattice speed $c$ is set to be 1 ($\delta_x = \delta_t$) in this work.

The MRT-LB equation with a semi-implicit treatment of the source term can be written as

$$f_i(\mathbf{x}+\mathbf{e}_i\delta_t, t+\delta_t) - f_i(\mathbf{x},t) = -\tilde{\Lambda}_{ij}(f_j - f_j^{eq})\Big|_{(\mathbf{x},t)} + \frac{\delta_t}{2}\left[S_i\Big|_{(\mathbf{x},t)} + S_i\Big|_{(\mathbf{x}+\mathbf{e}_i\delta_t, t+\delta_t)}\right], \tag{6}$$

where $f_i(\mathbf{x},t)$ is the density distribution function, $f_i^{eq}(\mathbf{x},t)$ is the equilibrium density distribution function, $S_i$ is the source term, and $\tilde{\mathbf{\Lambda}}$ is the collision matrix. Eq. (6) is implicit and cannot be directly implemented in simulations. By using a transformed distribution function $\bar{f}_i = f_i - 0.5\delta_t S_i$ [5], the explicit MRT-LB equation can be written in general as the following [23, 33, 34]:

$$\bar{f}_i\left(\mathbf{x}+\mathbf{e}_i\delta_t, t+\delta_t\right) = \bar{f}_i\left(\mathbf{x}, t\right) - \tilde{\Lambda}_{ij}\left(\bar{f}_j - f_j^{eq}\right)\Big|_{(\mathbf{x},t)} + \delta_t\left(S_i - 0.5\tilde{\Lambda}_{ij}S_j\right)\Big|_{(\mathbf{x},t)}. \tag{7}$$

The MRT-LB equation (7) consists of two processes: the collision process and streaming process. Through a transformation matrix $\mathbf{M}$, the collision process can be executed in the moment space:

$$\bar{\mathbf{m}}^*(\mathbf{x},t) = \bar{\mathbf{m}}(\mathbf{x},t) - \Lambda\left(\bar{\mathbf{m}} - \mathbf{m}^{eq}\right)\Big|_{(\mathbf{x},t)} + \delta_t\left(\mathbf{I} - \frac{\Lambda}{2}\right)\tilde{\mathbf{S}}, \tag{8}$$

while the streaming process is still executed in the velocity space:

$$\bar{f}_i\left(\mathbf{x}+\mathbf{e}_i\delta_t, t+\delta_t\right) = \bar{f}_i^*(\mathbf{x},t), \tag{9}$$

where $\Lambda = \mathbf{M}\tilde{\Lambda}\mathbf{M}^{-1} = \mathrm{diag}(s_0, s_1, \ldots, s_8)$ is a diagonal relaxation matrix ($\mathbf{M}^{-1}$ is the inverse matrix of $\mathbf{M}$ and $\{s_i\}$ are relaxation rates), $\mathbf{I}$ is the identity matrix, and the bold-face symbols $\mathbf{m}$, $\bar{\mathbf{m}}$, $\mathbf{m}^{eq}$ and $\tilde{\mathbf{S}}$ denote 9-dimensional column vectors of moments as follows:

$$\mathbf{m} = \mathbf{M}\mathbf{f} = |m\rangle, \quad \bar{\mathbf{m}} = \mathbf{M}\bar{\mathbf{f}} = |\bar{m}\rangle, \quad \mathbf{m}^{eq} = \mathbf{M}\mathbf{f}^{eq} = |m^{eq}\rangle, \quad \tilde{\mathbf{S}} = \mathbf{M}\mathbf{S} = |\tilde{S}\rangle, \tag{10}$$

in which $\mathbf{f} = |f\rangle$, $\bar{\mathbf{f}} = |\bar{f}\rangle$, $\mathbf{f}^{eq} = |f^{eq}\rangle$, and $\mathbf{S} = |S\rangle$. For brevity, the notation $|\cdot\rangle$ is employed to denote a 9-dimenaional column vector, e.g., $|m\rangle = (m_0, m_1, \ldots, m_8)^\mathrm{T}$. The post-collision distribution functions $\{\bar{f}_i^*\}$ can be determined by $\bar{\mathbf{f}}^* = \mathbf{M}^{-1}\bar{\mathbf{m}}^* = |\bar{f}^*\rangle$.

The transformation matrix $\mathbf{M}$ transforms a vector in the velocity space $\mathbb{V}$ (spanned by $|f\rangle$) into a vector in the moment space $\mathbb{M}$ (spanned by $|m\rangle$). In the present work, the transformation matrix $\mathbf{M}$ is constructed based on nine non-orthogonal basis vectors as

$$\mathbf{M} = \left[|M_0\rangle, |M_1\rangle, |M_2\rangle, |M_3\rangle, |M_4\rangle, |M_5\rangle, |M_6\rangle, |M_7\rangle, |M_8\rangle\right]^\mathrm{T}. \tag{11}$$

where the nine non-orthogonal basis vectors are obtained from the combinations of the lattice velocity components ($e_{ix}^p e_{iy}^q$, $p, q \in \{0, 1, 2\}$) in an ascending order as follows ($c = 1$) [30]:

$$|M_0\rangle = \left||\mathbf{e}_i|^0\right\rangle = (1,1,1,1,1,1,1,1,1)^\mathrm{T},$$

$$|M_1\rangle = |e_x\rangle = (0,1,0,-1,0,1,-1,-1,1)^\mathrm{T},$$

$$|M_2\rangle = |e_y\rangle = (0,0,1,0,-1,1,1,-1,-1)^\mathrm{T},$$

$$|M_3\rangle = |e_x^2 + e_y^2\rangle = (0,1,1,1,1,2,2,2,2)^{\mathrm{T}},$$

$$|M_4\rangle = |e_x^2 - e_y^2\rangle = (0,1,-1,1,-1,0,0,0,0)^{\mathrm{T}}, \tag{12}$$

$$|M_5\rangle = |e_x e_y\rangle = (0,0,0,0,0,1,-1,1,-1)^{\mathrm{T}},$$

$$|M_6\rangle = |e_x^2 e_y\rangle = (0,0,0,0,0,1,1,-1,-1)^{\mathrm{T}},$$

$$|M_7\rangle = |e_x e_y^2\rangle = (0,0,0,0,0,1,-1,-1,1)^{\mathrm{T}},$$

$$|M_8\rangle = |e_x^2 e_y^2\rangle = (0,0,0,0,0,1,1,1,1)^{\mathrm{T}}.$$

Explicitly, the non-orthogonal transformation matrix $\mathbf{M}$ can be expressed as

$$\mathbf{M} = \begin{pmatrix} 1 & 1 & 1 & 1 & 1 & 1 & 1 & 1 & 1 \\ 0 & 1 & 0 & -1 & 0 & 1 & -1 & -1 & 1 \\ 0 & 0 & 1 & 0 & -1 & 1 & 1 & -1 & -1 \\ 0 & 1 & 1 & 1 & 1 & 2 & 2 & 2 & 2 \\ 0 & 1 & -1 & 1 & -1 & 0 & 0 & 0 & 0 \\ 0 & 0 & 0 & 0 & 0 & 1 & -1 & 1 & -1 \\ 0 & 0 & 0 & 0 & 0 & 1 & 1 & -1 & -1 \\ 0 & 0 & 0 & 0 & 0 & 1 & -1 & -1 & 1 \\ 0 & 0 & 0 & 0 & 0 & 1 & 1 & 1 & 1 \end{pmatrix}. \tag{13}$$

With the transformation matrix $\mathbf{M}$ given above, the moments $\mathbf{m}$ and $\bar{\mathbf{m}}$ are defined as follows:

$$\mathbf{m} = |m\rangle = (\rho, j_x, j_y, e, p_{xx}, p_{xy}, q_{xxy}, q_{xyy}, \varepsilon)^{\mathrm{T}}, \tag{14}$$

$$\bar{\mathbf{m}} = |\bar{m}\rangle = (\rho, j_x - \tfrac{\delta_t}{2}\rho F_x, j_y - \tfrac{\delta_t}{2}\rho F_y, \bar{e}, \bar{p}_{xx}, \bar{p}_{xy}, \bar{q}_{xxy}, \bar{q}_{xyy}, \bar{\varepsilon})^{\mathrm{T}}, \tag{15}$$

where $j_x = \rho u_x$ and $j_y = \rho u_y$ are $x$- and $y$-components of the momentum $\rho \mathbf{u} = (j_x, j_y)$, and $\mathbf{m} = \bar{\mathbf{m}} + 0.5\delta_t \tilde{\mathbf{S}}$ ($\bar{f}_i = f_i - 0.5\delta_t S_i$). For $\mathbf{m}$, only the density $\rho$ and the two components of the momentum ($j_x$ and $j_y$) are conserved moments, while the other moments are non-conserved quantities. The equilibrium moments $\{m_i^{eq} | i = 0,1,\ldots,8\}$ are given by

$$m_0^{eq} = \rho, \quad m_1^{eq} = \rho u_x, \quad m_2^{eq} = \rho u_y,$$

$$m_3^{eq} = \frac{2}{3}\rho + \rho(u_x^2 + u_y^2), \quad m_4^{eq} = \rho(u_x^2 - u_y^2), \quad m_5^{eq} = \rho u_x u_y, \tag{16}$$

$$m_6^{eq} = \frac{1}{3}\rho u_y, \quad m_7^{eq} = \frac{1}{3}\rho u_x, \quad m_8^{eq} = \frac{1}{9}\rho + \frac{1}{3}\rho(u_x^2 + u_y^2).$$

The components of the source term $\tilde{\mathbf{S}}$ in the moment space are given as follows:

$$\tilde{S}_0 = 0, \quad \tilde{S}_1 = \rho F_x, \quad \tilde{S}_2 = \rho F_y, \quad \tilde{S}_3 = 2\rho(u_x F_x + u_y F_y), \quad \tilde{S}_4 = 2\rho(u_x F_x - u_y F_y),$$

$$\tilde{S}_5 = \rho(u_x F_y + u_y F_x), \quad \tilde{S}_6 = \frac{1}{3}\rho F_y, \quad \tilde{S}_7 = \frac{1}{3}\rho F_x, \quad \tilde{S}_8 = \frac{2}{3}\rho(u_x F_y + u_y F_x). \quad (17)$$

The macroscopic fluid density $\rho$ and velocity $\mathbf{u}$ can be calculated by

$$\rho = \sum_{i=0}^{8} f_i = \sum_{i=0}^{8} \bar{f}_i, \quad (18)$$

$$\rho \mathbf{u} = \sum_{i=0}^{8} \mathbf{e}_i f_i = \sum_{i=0}^{8} \mathbf{e}_i \bar{f}_i + \frac{\delta_t}{2}\rho \mathbf{F}. \quad (19)$$

The pressure $p$ is defined as $p = \rho c_s^2$, where $c_s$ is the lattice sound speed of the D2Q9 model and $c_s^2 = c^2/3 = 1/3$. For incompressible flows considered in this work, the so-called incompressible approximation [35] can be employed in the present MRT-LB model, i.e., the density $\rho = \rho_0 + \delta\rho \approx \rho_0$ ($\delta\rho$ is the density fluctuation), $j_x = \rho_0 u_x$, and $j_y = \rho_0 u_y$.

The diagonal relaxation matrix $\mathbf{\Lambda}$ is given by

$$\mathbf{\Lambda} = \mathrm{diag}(s_0, s_1, s_2, s_3, s_4, s_5, s_6, s_7, s_8).$$

$$= \mathrm{diag}(1, 1, 1, s_e, s_\upsilon, s_\upsilon, s_q, s_q, s_\varepsilon). \quad (20)$$

Through the Chapman-Enskog analysis [30, 36], the macroscopic governing equations (1) and (2) can be recovered from the MRT-LB model in the incompressible limit. The kinematic viscosity $\upsilon$ and bulk viscosity $\xi$ are defined by $\upsilon = c_s^2(s_\upsilon^{-1} - 0.5)\delta_t$ and $\xi = c_s^2(s_e^{-1} - 0.5)\delta_t$, respectively.

In the D2Q9 non-orthogonal MRT-LB model, the strain rate tensor $\mathbf{S} = [\nabla \mathbf{u} + (\nabla \mathbf{u})^\top]/2$ can be obtained from the non-equilibrium part of the velocity moments, i.e., $\bar{\mathbf{m}}^{neq} = \bar{\mathbf{m}}^{(1)} \approx \bar{\mathbf{m}} - \mathbf{m}^{eq}$. The components of the strain rate tensor can be expressed in terms of $\bar{m}_3^{(1)}$, $\bar{m}_4^{(1)}$, and $\bar{m}_5^{(1)}$, which are given as follows:

$$\bar{m}_3^{(1)} = -\frac{2}{3s_3}\rho\delta_t(\partial_x u_x + \partial_y u_y) - 0.5\delta_t \tilde{S}_3,$$

$$\bar{m}_4^{(1)} = -\frac{2}{3s_4}\rho\delta_t(\partial_x u_x - \partial_y u_y) - 0.5\delta_t \tilde{S}_4, \quad (21)$$

$$\bar{m}_5^{(1)} = -\frac{1}{3s_5}\rho\delta_t(\partial_x u_y + \partial_y u_x) - 0.5\delta_t \tilde{S}_5,$$

The equilibrium distribution function $f_i^{eq}$ ($\mathbf{f}^{eq} = \mathbf{M}^{-1}\mathbf{m}^{eq}$) in the velocity space is given by [16]

$$f_i^{eq} = \omega_i \rho \left[ 1 + \frac{\mathbf{e}_i \cdot \mathbf{u}}{c_s^2} + \frac{(\mathbf{e}_i \cdot \mathbf{u})^2}{2c_s^4} - \frac{|\mathbf{u}|^2}{2c_s^2} \right]. \tag{22}$$

The weight coefficient $\omega_i$ is given by $\omega_0 = 4/9$, $\omega_{1\sim 4} = 1/9$, and $\omega_{5\sim 8} = 1/36$.

## 2.2 MRT-LB model for the temperature field

In this subsection, a passive scalar thermal MRT-LB model is proposed for the temperature field. The MRT-LB equation can be expressed as

$$g_i(\mathbf{x} + \mathbf{e}_i \delta_t, t + \delta_t) - g_i(\mathbf{x}, t) = -(\mathbf{N}^{-1}\mathbf{Q}\mathbf{N})_{ij}(g_j - g_j^{eq})\Big|_{(\mathbf{x}, t)}, \tag{23}$$

where $g_i(\mathbf{x}, t)$ is the temperature distribution function, $g_i^{eq}(\mathbf{x}, t)$ is the equilibrium temperature distribution function, $\mathbf{Q}$ is a relaxation matrix, and $\mathbf{N}$ is a transformation matrix. Through the transformation matrix $\mathbf{N}$, the collision process of the MRT-LB equation (23) can be executed in the moment space:

$$\mathbf{n}^*(\mathbf{x}, t) = \mathbf{n}(\mathbf{x}, t) - \mathbf{Q}(\mathbf{n} - \mathbf{n}^{eq})\Big|_{(\mathbf{x}, t)}. \tag{24}$$

The streaming process is still executed in the velocity space:

$$g_i(\mathbf{x} + \mathbf{e}_i \delta_t, t + \delta_t) = g_i^*(\mathbf{x}, t). \tag{25}$$

Here, $\mathbf{n}$ and $\mathbf{n}^{eq}$ are vectors defined by $\mathbf{n} = \mathbf{N}\mathbf{g} = |n\rangle$ and $\mathbf{n}^{eq} = \mathbf{N}\mathbf{g}^{eq} = |n^{eq}\rangle$, respectively. The post-collision distribution functions $\{g_i^*\}$ are determined by $\mathbf{g}^* = \mathbf{N}^{-1}\mathbf{n}^* = |g^*\rangle$ ($\mathbf{N}^{-1}$ is the inverse matrix of $\mathbf{N}$ given in Appendix A).

The transformation matrix $\mathbf{N}$ projects a vector (in the velocity space) onto the moment space. For the temperature field, the D2Q5 lattice is employed (the five discrete velocities $\{\mathbf{e}_i | i = 0, 1, \ldots, 4\}$ are given in Eq. (5)), and the non-orthogonal transformation matrix $\mathbf{N}$ can be defined as ($c = 1$)

$$\mathbf{N} = \left[ |1\rangle, |e_x\rangle, |e_y\rangle, |e_x^2 + e_y^2\rangle, |e_x^2 - e_y^2\rangle \right]^T$$

$$= \begin{pmatrix} 1 & 1 & 1 & 1 & 1 \\ 0 & 1 & 0 & -1 & 0 \\ 0 & 0 & 1 & 0 & -1 \\ 0 & 1 & 1 & 1 & 1 \\ 0 & 1 & -1 & 1 & -1 \end{pmatrix}. \tag{26}$$

In the model, $n_0$ is the only conserved moment and the temperature $T$ is computed by

$$T \equiv n_0 = \sum_{i=0}^{4} g_i. \tag{27}$$

The equilibrium moments $\{n_i^{eq} | i = 0,1,\ldots,4\}$ for the moments $\{n_i | i = 0,1,\ldots,4\}$ are defined as

$$n_0^{eq} = T, \quad n_1^{eq} = u_x T, \quad n_2^{eq} = u_y T, \quad n_3^{eq} = \varpi T, \quad n_4^{eq} = 0, \tag{28}$$

where $\varpi \in (0,1)$ is a parameter of the model.

For incompressible thermal flows with isotropic thermal diffusivity, the relaxation matrix $\mathbf{Q}$ is a diagonal one and is given by

$$\mathbf{Q} = \text{diag}(\zeta_0, \zeta_1, \zeta_2, \zeta_3, \zeta_4).$$

$$= \text{diag}(1, \zeta_\alpha, \zeta_\alpha, \zeta_e, \zeta_\upsilon). \tag{29}$$

Through the Chapman-Enskog analysis of the MRT-LB equation (23) in the moment space (see Appendix B for details), the temperature equation (3) can be recovered. The thermal diffusivity $\alpha$ is defined as $\alpha = c_{sT}^2 (\zeta_\alpha^{-1} - 0.5)\delta_t$, in which $c_{sT}^2 = \varpi/2$ ($c_{sT}$ is the lattice sound speed of the D2Q5 model). The equilibrium temperature distribution function $g_i^{eq}$ ($\mathbf{g}^{eq} = \mathbf{N}^{-1}\mathbf{n}^{eq}$) in the velocity space is given by

$$g_i^{eq} = \begin{cases} (1-\varpi)T, & i = 0, \\ \dfrac{1}{4}\varpi T + \dfrac{1}{2}(\mathbf{e}_i \cdot \mathbf{u})T, & i = 1 \sim 4. \end{cases} \tag{30}$$

We would like to point out that the thermal MRT-LB model for the temperature field can also be developed based on the D2Q9 lattice. In what follows, the thermal D2Q9 MRT-LB model is briefly introduced. The equilibrium moments $\{n_i^{eq} | i = 0,1,\ldots,8\}$ can be obtained via $\mathbf{n}^{eq} = \mathbf{N}\mathbf{g}^{eq}$, in which the equilibrium temperature distribution function $g_i^{eq}$ is given by

$$f_i^{eq} = \omega_i T \left\{ 1 + \frac{\mathbf{e}_i \cdot \mathbf{u}}{c_s^2} + \vartheta \left[ \frac{(\mathbf{e}_i \cdot \mathbf{u})^2}{2c_s^4} - \frac{|\mathbf{u}|^2}{2c_s^2} \right] \right\}, \tag{31}$$

where $\vartheta \in \{0,1\}$. The diagonal relaxation matrix $\mathbf{Q}$ is given by

$$\mathbf{Q} = \mathrm{diag}\left(1, \zeta_\alpha, \zeta_\alpha, \zeta_e, \zeta_\upsilon, \zeta_\upsilon, \zeta_q, \zeta_q, \zeta_\varepsilon \right). \tag{32}$$

The thermal diffusivity $\alpha$ is defined as $\alpha = c_s^2 \left( \zeta_\alpha^{-1} - 0.5 \right) \delta_t$. Through the Chapman-Enskog analysis, the following macroscopic equation can be obtained

$$\partial_t T + \nabla \cdot (\mathbf{u}T) = \nabla \cdot \left\{ \alpha \nabla T + \delta_t \left( \zeta_\alpha^{-1} - 0.5 \right) \left[ \epsilon \partial_{t_1}(\mathbf{u}T) + \vartheta \nabla \cdot (\mathbf{uu}T) \right] \right\}. \tag{33}$$

As compared with the temperature equation (3), Eq. (33) contains a deviation term $\nabla \cdot \left\{ \delta_t \left( \zeta_\alpha^{-1} - 0.5 \right) \left[ \epsilon \partial_{t_1}(\mathbf{u}T) + \vartheta \nabla \cdot (\mathbf{uu}T) \right] \right\}$. In most cases, this deviation term can be ignored for incompressible thermal flows.

### 3. Numerical simulations

In this section, numerical simulations of several typical 2D problems are carried out to validate the effectiveness and accuracy of the MRT-LB method. In the present study, we set $\rho_0 = 1$, $\delta_x = \delta_y = \delta_t = 1$ ($c = 1$), and $\varpi = 1/2$ ($c_{sT}^2 = 1/4$). Unless otherwise specified, the non-equilibrium extrapolation scheme [37] is employed to treat the velocity and temperature boundary conditions in simulations.

*3.1 Fully developed channel flow driven by a constant body force*

We first consider the fully developed flow in a channel between two parallel plates driven by a constant body force to validate the accuracy of the D2Q9 MRT-LB model presented in Section 2.1. The length and width of the channel are $L$ and $H$, respectively. A constant body force $\mathbf{a} = (a_x, 0)$ is imposed in the flow direction (*x*-direction). The analytical solution of this problem is given by [38]

$$u_x(y) = u_0 \left[ 1 - (y/l_c)^2 \right], \tag{34}$$

where $0 \leq x \leq L$, $-H/2 \leq y \leq H/2$, and $u_0 = a_x l_c^2 / (2\upsilon)$ is the maximum velocity ($l_c$ is the characteristic length and $l_c = H/2$ ). The Reynolds number is defined by $Re = u_0 l_c / \upsilon$. The non-equilibrium extrapolation scheme is employed to treat the no-slip velocity boundary condition at the upper and bottom plates, and periodic boundary conditions are imposed at the inlet and outlet of the channel. The relaxation rates $s_4$ and $s_5$ ( $s_{4,5} = s_\upsilon$ ) are set to be $1.754$, while the remaining relaxation rates are set to be $1.0$ in our simulations. A grid size of $N_x \times N_y = 6 \times 60$ is adopted to resolve the computational domain.

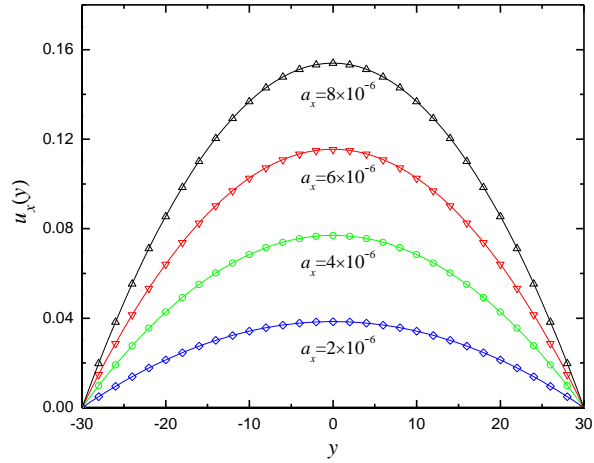

Fig. 1. Velocity profiles for different values of body force $a_x$ (symbols represent MRT-LB results and lines represent analytical solutions).

The profiles of the non-normalized horizontal velocity $u_x(y)$ for different values of body force $a_x$ are plotted in Fig. 1. It can be seen from Fig. 1 that the numerical results agree well with the analytical solutions. To quantify the difference between the numerical and analytical solutions, the relative global errors of velocity for different values of body force $a_x$ are measured. The relative global error of velocity is calculated by

$$E_{\mathbf{u}} = \frac{\sum_{\mathbf{x}} \|\mathbf{u}_A(\mathbf{x}) - \mathbf{u}_{LB}(\mathbf{x})\|_2}{\sum_{\mathbf{x}} \|\mathbf{u}_A(\mathbf{x})\|_2}, \tag{35}$$

where $\mathbf{u}_A$ and $\mathbf{u}_{LB}$ are the analytical and numerical solutions, respectively, and $\|\cdot\|_2$ denotes the L2-norm. The relative global errors for different values of body force $a_x$ are listed in Tab. 1. As shown in Tab. 1, the relative global errors are rather small, and for the parameters and resolution given above, the relative global error $E_{\mathbf{u}} \sim O(10^{-3})$.

Tab. 1. Relative global errors of velocity for different values of body force $a_x$.

| Body force ($a_x$) | Reynolds number ($Re$) | Relative global error ($E_{\mathbf{u}}$) |
|---|---|---|
| $2 \times 10^{-6}$ | 49.4146 | $1.2600 \times 10^{-3}$ |
| $4 \times 10^{-6}$ | 98.8292 | $1.2585 \times 10^{-3}$ |
| $6 \times 10^{-6}$ | 148.244 | $1.2560 \times 10^{-3}$ |
| $8 \times 10^{-6}$ | 197.658 | $1.2524 \times 10^{-3}$ |

In what follows, numerical simulations are carried out to evaluate the spatial accuracy of the present MRT-LB model for different parameters at $Re = 49.4146$. Three different values of $s_\upsilon$ are considered ($1.60, 1.754$, and $1.90$), and for each case, the lattice spacing $\delta_x = 1/N_y$ varies from $1/30$ to $1/150$. The relative global errors of velocity for different relaxation rates and lattice spacings are computed and listed in Tab. 2. As shown in Tab. 2, for a given lattice spacing, the relative global errors at different relaxation rates are comparable to each other. Furthermore, the relative global errors of velocity for different relaxation rates and lattice spacings are plotted logarithmically in Fig. 2. The slopes of the fitting lines in Fig. 2 are about $1.9889$, $1.9902$, and $1.9912$ for $s_\upsilon = 1.60$, $1.754$, and $1.90$, respectively. These results indicate that the D2Q9 MRT-LB model is of second-order accuracy in space.

Tab. 2. Relative global errors of velocity for fully developed flow in a channel driven by a constant body force at $Re = 49.4146$.

| Relaxation rate ($s_\upsilon$) | Relative global error ($E_{\mathbf{u}}$) | | | | |
|---|---|---|---|---|---|
| | $\delta_x = 1/30$ | $\delta_x = 1/60$ | $\delta_x = 1/90$ | $\delta_x = 1/120$ | $\delta_x = 1/150$ |
| 1.60 | $4.2709 \times 10^{-3}$ | $1.0804 \times 10^{-3}$ | $4.8184 \times 10^{-4}$ | $2.7148 \times 10^{-4}$ | $1.7391 \times 10^{-4}$ |
| 1.754 | $4.9910 \times 10^{-3}$ | $1.2600 \times 10^{-3}$ | $5.6177 \times 10^{-4}$ | $3.1674 \times 10^{-4}$ | $2.0270 \times 10^{-4}$ |
| 1.90 | $5.5980 \times 10^{-3}$ | $1.4119 \times 10^{-3}$ | $6.2932 \times 10^{-4}$ | $3.5450 \times 10^{-4}$ | $2.2715 \times 10^{-4}$ |

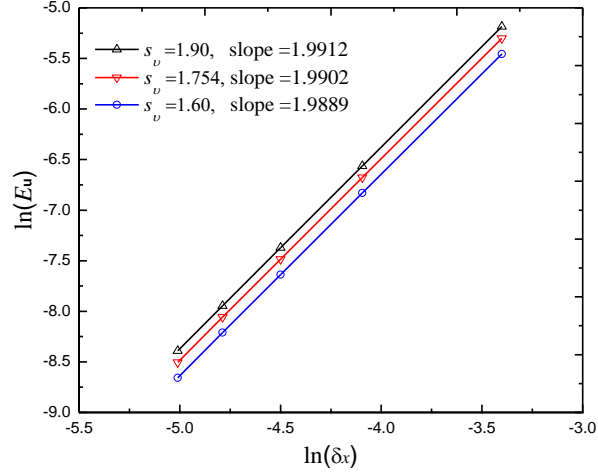

Fig. 2. Relative global error of velocity versus lattice spacing for fully developed flow in a channel driven by a constant body force at $Re = 49.4146$.

### 3.2 Lid-driven cavity flow

The 2D lid-driven cavity flow is a standard benchmark problem for testing numerical schemes. The top wall of the cavity moves from left to right (*x*-direction) with a uniform velocity $u_0$, while the left, bottom, and right walls are fixed. This flow is characterized by the Reynolds number $Re = Lu_0/\upsilon$, where $L$ is the length of the cavity, and $H$ is the height of the cavity ($H/L = 1$). The D2Q9 MRT-LB model is used to study the flow at $Re = 1000$. In simulations, the driven velocity $u_0$ is set to be 0.1, and a $192\times192$ uniform mesh is adopted. The relaxation rate $s_\upsilon$ is determined by $s_\upsilon^{-1} = 0.5 + Lu_0/(c_s^2 \delta_t Re)$, and the remaining relaxation rates are chosen as follows: $s_0 = s_1 = s_2 = 1$, $s_3 = 1.6$, $s_6 = s_7 = s_q = 1.2$, and $s_8 = 1.8$. The horizontal velocity component $u_x$ at the vertical midplane ($x/L = 0.5$) and the vertical velocity component $u_y$ at the horizontal midplane ($y/H = 0.5$) are shown in Fig. 3. The benchmark solutions of Ghia et al. [39] are also included in the figure for comparison. Clearly, the present results are in good agreement with the benchmark data.

The comparison of the numerical stability between the BGK model and the non-orthogonal MRT model is made at $Re = 1200$ on a $80\times80$ uniform mesh. The streamlines and the contour lines of the velocity field predicted by both models are shown in Figs. 4 and 5, respectively. It is found that the streamlines and the contour lines of the velocity field predicted by the BGK model at $Re = 1200$ on a $80 \times 80$ uniform mesh have unphysical oscillations, while those predicted by the non-orthogonal MRT model are still smooth. These results indicate that the non-orthogonal MRT model has a better numerical stability than the BGK model with the same boundary treatment. Furthermore, it is found that the computation time of the non-orthogonal MRT model is about 4% less than that of the classical orthogonal MRT model [19].

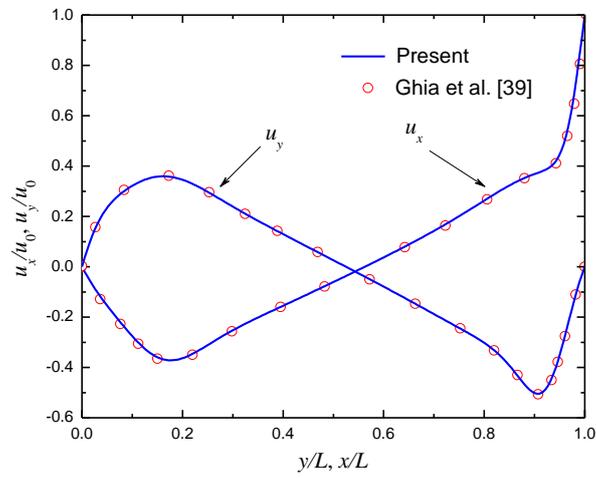

Fig. 3. Velocity profiles through the center of the cavity at $Re = 1000$ on a $192 \times 192$ uniform mesh.

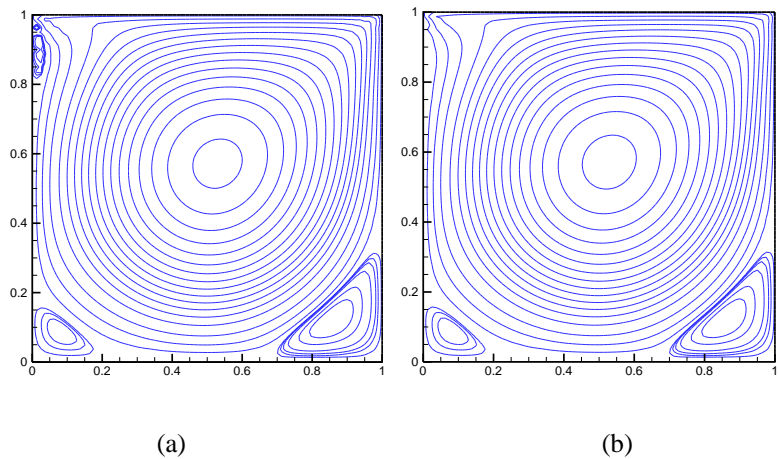

(a)  (b)

Fig. 4. Streamlines of the cavity flow at $Re = 1200$ on a $80 \times 80$ uniform mesh: (a) BGK model and (b) present non-orthogonal MRT model.

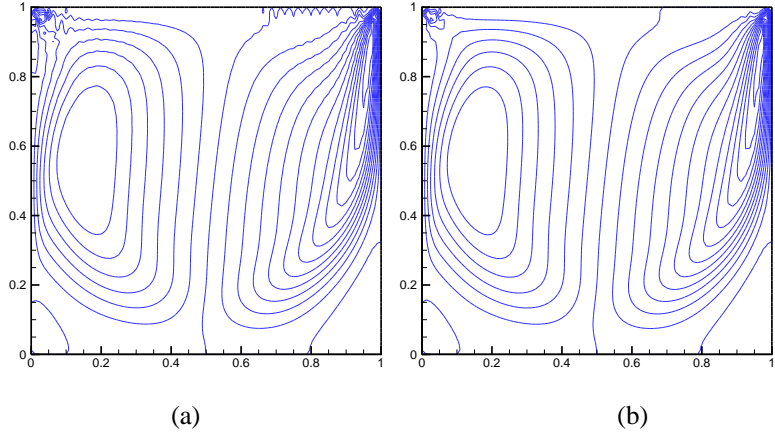

(a)                         (b)

Fig. 5. Contour lines of the velocity field at $Re=1200$ on a $80\times 80$ uniform mesh: (a) BGK model and (b) present non-orthogonal MRT model.

### 3.3 Thermal flow in a channel with wall injection

In this subsection, the MRT-LB method is applied to simulate the fully developed thermal flow in a channel [6, 40], where the upper hot plate ($T=T_h$) moves along the $x$-direction with a uniform velocity $u_0$, and a constant normal flow of fluid is injected (with velocity $v_0$) through the static bottom cold plate ($T=T_c$) and withdrawn from the upper plate at the same rate. In steady state, the analytical solution of the velocity field is given by [38]

$$u_x(y) = u_0 \frac{\exp(Re\cdot y/H)-1}{\exp(Re)-1}, \tag{36}$$

where $Re$ is the Reynolds number defined by $Re=Hv_0/\upsilon$, and $H$ is the width of the channel. The temperature profile in steady state is given by [6, 40]

$$T = T_c + \Delta T \frac{\exp(PrRe\cdot y/H)-1}{\exp(PrRe)-1}, \tag{37}$$

where $\Delta T = T_h - T_c$ is the temperature difference.

In simulations, we set $Pr=0.71$, $u_0=v_0=0.01$, $T_h=1$, and $T_c=0$. The non-equilibrium extrapolation scheme is employed to the upper and bottom plates for the velocity and temperature boundary conditions, and periodic boundary conditions are imposed at the inlet and outlet of the channel. A grid size of $N_x \times N_y = 30 \times 60$ is employed, and the relaxation rates $s_\upsilon$ and $\zeta_\alpha$ are determined by $s_\upsilon^{-1} = 0.5 + Hv_0 / (c_s^2 \delta_t Re)$ and $\zeta_\alpha^{-1} = 0.5 + c_s^2 (s_\upsilon^{-1} - 0.5)/(c_{sT}^2 Pr)$, respectively. The remaining relaxation rates are set to be $1.0$ in simulations. The velocity and temperature profiles for different Reynolds numbers at $Pr = 0.71$ are plotted in Fig. 6. It is found that our results are in excellent agreement with the analytical solutions.

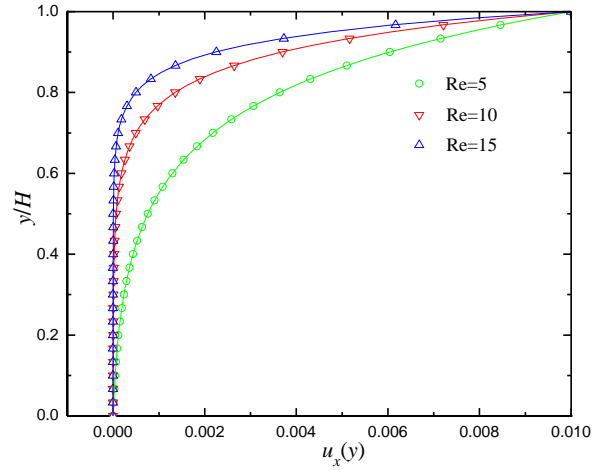

(a)

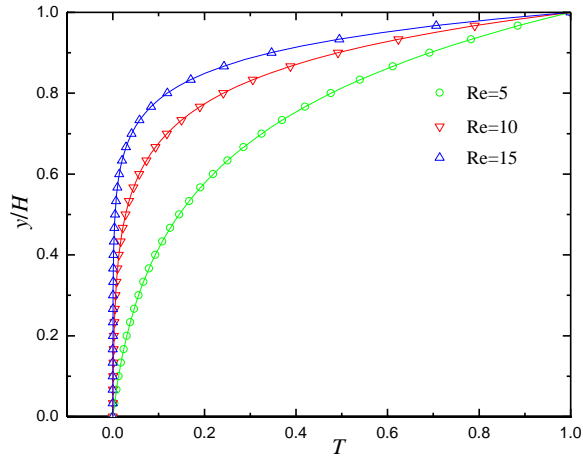

(b)

Fig. 6. Velocity (a) and temperature (b) profiles of thermal flow in a channel with wall injection for different Reynolds numbers at $Pr = 0.71$ (symbols represent MRT-LB results and lines represent analytical solutions).

In what follows, numerical simulations are carried out to evaluate the convergence rate of the present MRT-LB method. In the simulations, the Reynolds number $Re$ and relaxation rate $s_\upsilon$ are fixed at $10$ and $0.8$, respectively. The lattice spacing $\delta_x = 1/N_y$ varies from $1/30$ to $1/150$. The relative global error of temperature ($T$) is defined by

$$E_T = \frac{\sqrt{\sum_{\mathbf{x}}|T_A(\mathbf{x}) - T_{LB}(\mathbf{x})|^2}}{\sqrt{\sum_{\mathbf{x}}|T_A(\mathbf{x})|^2}}, \tag{38}$$

where $T_A$ and $T_{LB}$ are the analytical and numerical solutions, respectively, and the summation is over the entire domain. The relative global error of velocity is calculated by Eq. (35). The relative global errors of velocity and temperature for different lattice spacings are plotted logarithmically in Fig. 7. The slopes of the fitting lines in Fig. 7(a) and Fig. 7(b) are about $1.9670$ and $1.9948$, respectively. The results indicate that the present MRT-LB method has a second-order convergence rate in space.

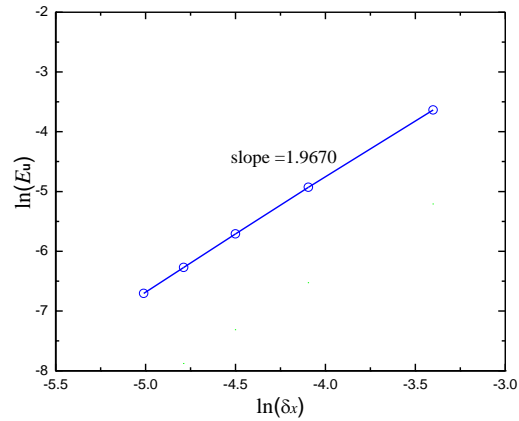

(a)

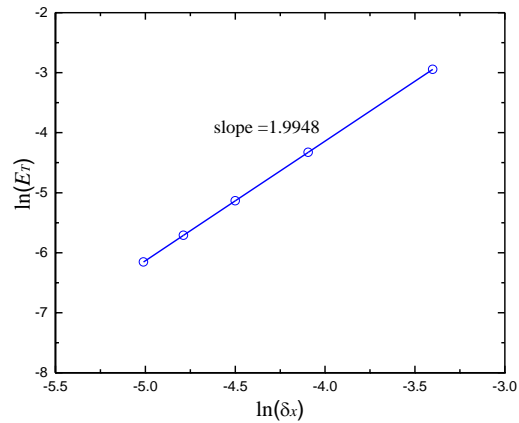

(a)

Fig. 7. Relative global errors of the velocity $\mathbf{u}$ (a) and temperature $T$ (b) versus lattice spacings at

$$Pr = 0.71 \text{ and } Re = 10.$$

*3.4 Natural convection in a square cavity*

In this subsection, the present MRT-LB method is employed to simulate natural convection flow in a square cavity. The horizontal walls of the cavity are adiabatic, while the left and right walls are kept at constant temperatures $T_h$ and $T_c$, respectively ($T_h > T_c$). The width and height of the cavity are $L$ and $H$ (for square cavity $H/L = 1$), respectively. The flow is characterized by the Prandtl number $Pr = \upsilon/\alpha$ and Rayleigh number $Ra = g\beta\Delta T L^3/(\upsilon\alpha)$, where $\Delta T = T_h - T_c$ is the temperature difference (characteristic temperature). The average Nusselt number $\overline{Nu}$ of the left (or right) wall is defined by

$$\overline{Nu} = \int_0^H Nu(y)\,dy/H, \tag{39}$$

where $Nu(y) = -L(\partial T/\partial x)_{\text{wall}}/\Delta T$ is the local Nusselt number. According to Ref. [41], the relaxation rates $s_\upsilon$ and $\zeta_\alpha$ can be determined by

$$s_\upsilon^{-1} = \frac{1}{2} + \frac{MaL\sqrt{3Pr}}{c\delta_t\sqrt{Ra}}, \quad \zeta_\alpha^{-1} = \frac{1}{2} + \frac{c_s^2\left(s_\upsilon^{-1} - 0.5\right)}{c_{sT}^2 Pr}, \tag{40}$$

respectively, where $Ma = u_c/c_s = \sqrt{3}u_c$ is the Mach number ($u_c = \sqrt{g\beta\Delta T L}$ is the characteristic velocity). The remaining relaxation parameters are chosen as follows: $s_0 = s_1 = s_2 = 1$, $s_3 = 1.6$, $s_6 = s_7 = s_q = 1.2$, $s_8 = 1.8$, $\zeta_0 = 1$, and $\zeta_3 = \zeta_4 = 1.9$.

In simulations, we set $Pr = 0.71$, $T_h = 21$, $T_c = 1$, $T_0 = (T_h + T_c)/2 = 11$, and $Ma = 0.1$. Through the grid-dependence study, the grid sizes of $128 \times 128$ for $Ra = 10^3$, $192 \times 192$ for $Ra = 10^4$, $256 \times 256$ for $Ra = 10^5$, and $256 \times 256$ for $10^6$ are employed to resolve the computational domain. The streamlines and isotherms for $Ra = 10^3 \sim 10^6$ are shown in Figs. 8 and 9, respectively. The temperature profiles at the horizontal midplane of the square cavity ($y/H = 0.5$) for $Ra = 10^3 \sim 10^6$ are plotted in Fig. 10. All the results agree well with those reported in previous works

[6, 40-43]. To quantify the results, the following quantities are computed: the maximum horizontal velocity component $u_{x,\max}$ in the vertical midplane $x = L/2$ and its location $y_{\max}$; the maximum vertical velocity component $u_{y,\max}$ in the horizontal midplane $y = H/2$ and its location $x_{\max}$; the average Nusselt number $\overline{Nu}$ along the right wall (cold wall), its maximum value $Nu_{\max}$ and the location $y_{Nu}$ where it occurs. The results predicted by the present MRT-LB method are compared with the results of previous works [6, 40, 42, 43] in Table 3. The present results are found to be in good agreement with the LB results [6, 40] and benchmark solutions [42, 43] reported in previous works.

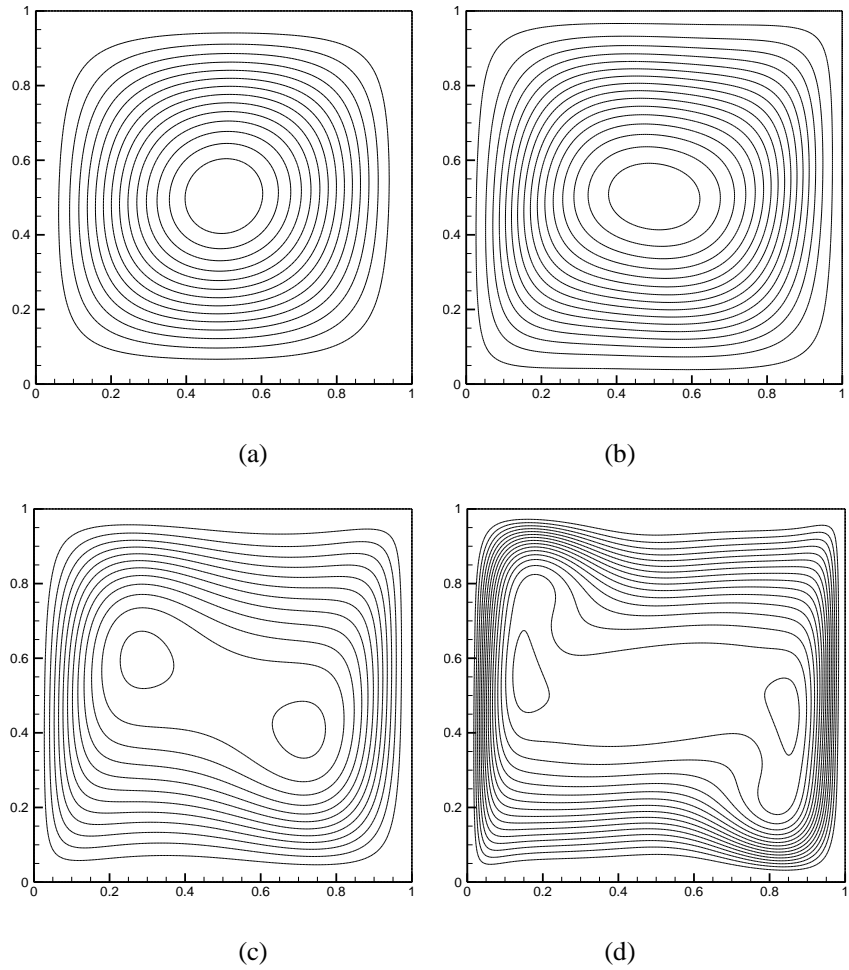

Fig. 8. Streamlines of natural convection flow in a square cavity: (a) $Ra = 10^3$, (b) $Ra = 10^4$, (c) $Ra = 10^5$ and (d) $Ra = 10^6$.

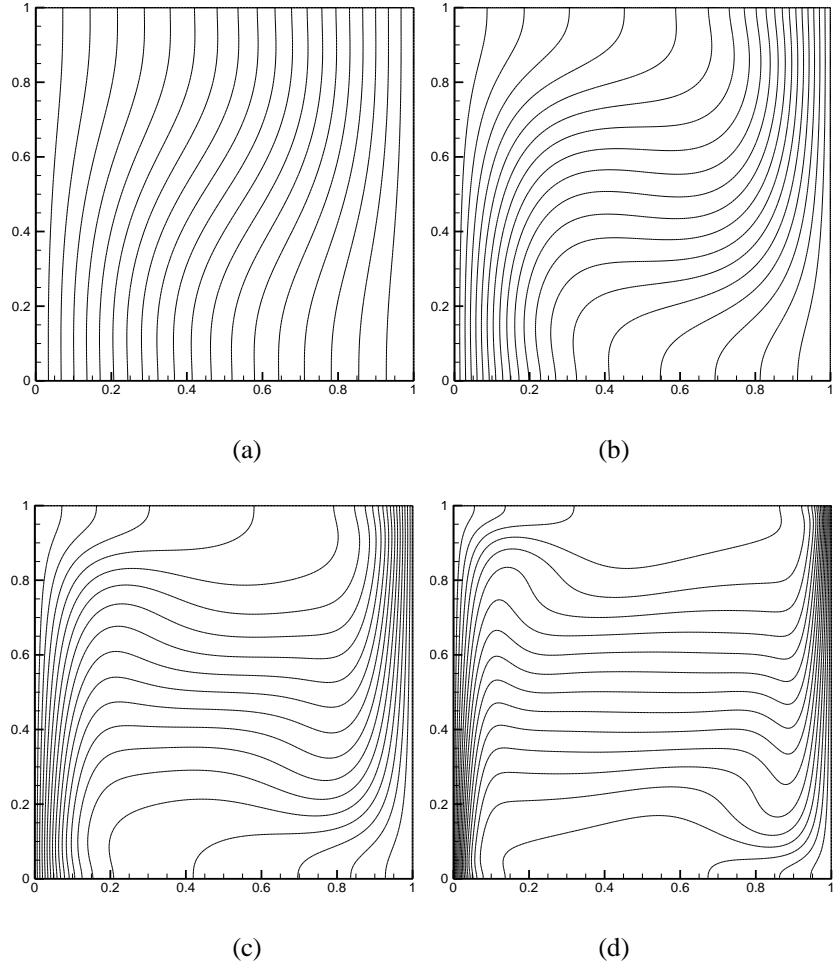

Fig. 9. Isotherms of natural convection flow in a square cavity: (a) $Ra = 10^3$, (b) $Ra = 10^4$, (c) $Ra = 10^5$ and (d) $Ra = 10^6$.

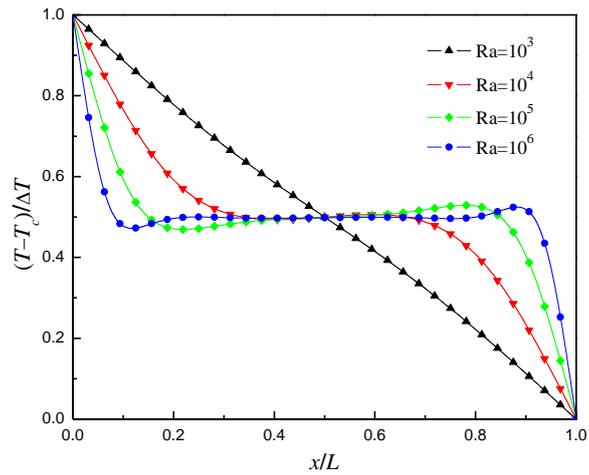

Fig. 10. Temperature profiles at the horizontal midplane of the square cavity ($y/H = 0.5$) for $Ra = 10^3 \sim 10^6$.

Tab. 3. Comparisons of the present results with the LB results [6, 40] and benchmark solutions [42, 43] for $Ra = 10^3 \sim 10^6$.

| $Ra$ | | $u_{x,\max}$ | $y_{\max}$ | $u_{y,\max}$ | $x_{\max}$ | $Nu_{\max}$ | $y_{Nu}$ | $\overline{Nu}$ |
|---|---|---|---|---|---|---|---|---|
| $10^3$ | LB [6] | 3.650 | 0.810 | 3.704 | 0.180 | - | - | 1.117 |
| | LB [40] | 3.6493 | 0.8125 | 3.7010 | 0.1797 | 1.5017 | 0.9063 | 1.1169 |
| | FDM [42] | 3.649 | 0.813 | 3.697 | 0.178 | - | - | - |
| | Present | 3.6528 | 0.8125 | 3.7004 | 0.1797 | 1.5005 | 0.9141 | 1.1161 |
| $10^4$ | LB [6] | 16.146 | 0.820 | 19.593 | 0.120 | - | - | 2.241 |
| | LB [40] | 16.1722 | 0.8229 | 19.6098 | 0.1198 | 3.5351 | 0.8594 | 2.2452 |
| | FVM [43] | 16.1802 | 0.8265 | 19.6295 | 0.1193 | 3.5309 | 0.8531 | 2.2448 |
| | Present | 16.177 | 0.8229 | 19.6184 | 0.1198 | 3.5301 | 0.8542 | 2.2447 |
| $10^5$ | LB [6] | 34.315 | 0.855 | 68.012 | 0.065 | - | - | 4.508 |
| | LB [40] | 34.7362 | 0.8555 | 68.4954 | 0.0664 | 7.7500 | 0.9219 | 4.5219 |
| | FVM [43] | 34.7399 | 0.8558 | 68.6396 | 0.0657 | 7.7201 | 0.9180 | 4.5216 |
| | Present | 34.6891 | 0.8555 | 68.5034 | 0.0664 | 7.7480 | 0.9219 | 4.5273 |
| $10^6$ | LB [6] | 63.671 | 0.852 | 217.57 | 0.040 | - | - | 8.737 |
| | LB [40] | 64.8687 | 0.8516 | 219.334 | 0.0391 | 17.6140 | 0.9648 | 8.7926 |
| | FVM [43] | 64.8367 | 0.8505 | 220.461 | 0.0390 | 17.5360 | 0.9608 | 8.8251 |
| | Present | 64.6968 | 0.8516 | 219.149 | 0.0391 | 17.6722 | 0.9648 | 8.8258 |

## 4. Conclusions

In this paper, a non-orthogonal MRT-LB method for simulating incompressible thermal flows is developed in the framework of the DDF approach. In the method, the incompressible N-S equations and temperature equation are solved separately by two different non-orthogonal MRT-LB equations, which are developed based on non-orthogonal basis vectors obtained from the combinations of the lattice velocity components. The present non-orthogonal MRT-LB method retains the computational efficiency and simplicity of the BGK-LB method. Numerical simulations of the fully developed channel flow driven by a constant body force, the lid-driven cavity flow, the thermal flow in a channel with wall injection, and the natural convection flow in a square cavity are carried out to validate the effectiveness and accuracy of the MRT-LB method. It is found that the present results are in good agreement with the analytical solutions and/or other numerical results reported in the literature. Numerical simulations of the lid-driven cavity flow indicate that the non-orthogonal MRT model has a better numerical stability than the BGK model, and the computation time of the non-orthogonal MRT model is less than that of the classical orthogonal MRT model. Furthermore, the grid convergence tests indicate that the present MRT-LB method has a second-order convergence rate in space. The extensions of the present method to simulate three-dimensional incompressible thermal flows will be considered in our future studies.

**Acknowledgements**

This work is supported by the Key Project of National Natural Science Foundation of China (No. 51436007) and the National Key Basic Research Program of China (973 Program) (2013CB228304).

**Appendix A: The inverse matrices of N and M**

$$\mathbf{M}^{-1} = \begin{pmatrix} 1 & 0 & 0 & -1 & 0 & 0 & 0 & 0 & 1 \\ 0 & \frac{1}{2} & 0 & \frac{1}{4} & \frac{1}{4} & 0 & 0 & -\frac{1}{2} & -\frac{1}{2} \\ 0 & 0 & \frac{1}{2} & \frac{1}{4} & -\frac{1}{4} & 0 & -\frac{1}{2} & 0 & -\frac{1}{2} \\ 0 & -\frac{1}{2} & 0 & \frac{1}{4} & \frac{1}{4} & 0 & 0 & \frac{1}{2} & -\frac{1}{2} \\ 0 & 0 & -\frac{1}{2} & \frac{1}{4} & -\frac{1}{4} & 0 & \frac{1}{2} & 0 & -\frac{1}{2} \\ 0 & 0 & 0 & 0 & 0 & \frac{1}{4} & \frac{1}{4} & \frac{1}{4} & \frac{1}{4} \\ 0 & 0 & 0 & 0 & 0 & -\frac{1}{4} & \frac{1}{4} & -\frac{1}{4} & \frac{1}{4} \\ 0 & 0 & 0 & 0 & 0 & \frac{1}{4} & -\frac{1}{4} & -\frac{1}{4} & \frac{1}{4} \\ 0 & 0 & 0 & 0 & 0 & -\frac{1}{4} & -\frac{1}{4} & \frac{1}{4} & \frac{1}{4} \end{pmatrix}, \qquad (A1)$$

$$\mathbf{N}^{-1} = \begin{pmatrix} 1 & 0 & 0 & -1 & 0 \\ 0 & \frac{1}{2} & 0 & \frac{1}{4} & \frac{1}{4} \\ 0 & 0 & \frac{1}{2} & \frac{1}{4} & -\frac{1}{4} \\ 0 & -\frac{1}{2} & 0 & \frac{1}{4} & \frac{1}{4} \\ 0 & 0 & -\frac{1}{2} & \frac{1}{4} & -\frac{1}{4} \end{pmatrix}. \qquad (A2)$$

**Appendix B: Chapman-Enskog analysis of the D2Q5 MRT-LB model**

The Chapman-Enskog expansion approach [36] is adopted to derive the temperature equation (3) from the D2Q5 MRT-LB equation (23) in the moment space. To this end, the following expansions in time and space are introduced [35]:

$$g_i\left(\mathbf{x}+\mathbf{e}_i\delta_t,\ t+\delta_t\right)=\sum_{i=0}^{\infty}\frac{\epsilon^n}{n!}\left(\partial_t+\mathbf{e}_i\cdot\nabla\right)^n g_i\left(\mathbf{x},\ t\right), \quad \text{(B1)}$$

$$g_i = g_i^{(0)} + \epsilon g_i^{(1)} + \epsilon^2 g_i^{(2)} + \cdots, \quad \text{(B2)}$$

$$\partial_t = \epsilon \partial_{t_1} + \epsilon^2 \partial_{t_2},\ \partial_j = \epsilon \partial_{j1}, \quad \text{(B3)}$$

where $\epsilon$ is a small expansion parameter. With the above expansions, we can derive the following equations from Eq. (23) as consecutive orders of the parameter $\epsilon$:

$$\epsilon^0:\ g_i^{(0)} = g_i^{eq}, \quad \text{(B4)}$$

$$\epsilon^1:\ D_{1i}g_i^{(0)} = -\frac{1}{\delta_t}\left(\mathbf{N}^{-1}\mathbf{Q}\mathbf{N}\right)_{ij} g_j^{(1)}, \quad \text{(B5)}$$

$$\epsilon^2:\ \partial_{t_2} g_i^{(0)} + D_{1i} g_i^{(1)} + \frac{\delta_t}{2} D_{1i}^2 g_i^{(0)} = -\frac{1}{\delta_t}\left(\mathbf{N}^{-1}\mathbf{Q}\mathbf{N}\right)_{ij} g_j^{(2)}, \quad \text{(B6)}$$

where $D_{1i} = \partial_{t_1} + \mathbf{e}_i \cdot \nabla_1 = \partial_{t_1} + e_{ij}\partial_{j1}$ ($j=x,y$). The above equations can be transformed into the moment space as follows:

$$\epsilon^0:\ \mathbf{n}^{(0)} = \mathbf{n}^{eq}, \quad \text{(B7)}$$

$$\epsilon^1:\ \left(\mathbf{I}\partial_{t_1} + \mathbf{E}\cdot\nabla_1\right)\mathbf{n}^{(0)} = -\frac{\mathbf{Q}}{\delta_t}\mathbf{n}^{(1)}, \quad \text{(B8)}$$

$$\epsilon^2:\ \partial_{t_2}\mathbf{n}^{(0)} + \left(\mathbf{I}\partial_{t_1} + \mathbf{E}\cdot\nabla_1\right)\left(\mathbf{I} - \frac{\mathbf{Q}}{2}\right)\mathbf{n}^{(1)} = -\frac{\mathbf{Q}}{\delta_t}\mathbf{n}^{(2)}, \quad \text{(B9)}$$

where $\mathbf{E} = \left(\mathbf{E}_x, \mathbf{E}_y\right)$, in which $\mathbf{E}_j = \mathbf{N}\left[\text{diag}(e_{0j}, e_{1j}, \ldots, e_{4j})\right]\mathbf{N}^{-1}$ ($j=x,y$) are given explicitly by

$$\mathbf{E}_x = \begin{pmatrix} 0 & 1 & 0 & 0 & 0 \\ 0 & 0 & 0 & \frac{1}{2} & \frac{1}{2} \\ 0 & 0 & 0 & 0 & 0 \\ 0 & 1 & 0 & 0 & 0 \\ 0 & 1 & 0 & 0 & 0 \end{pmatrix},\ \mathbf{E}_y = \begin{pmatrix} 0 & 0 & 1 & 0 & 0 \\ 0 & 0 & 0 & 0 & 0 \\ 0 & 0 & 0 & \frac{1}{2} & -\frac{1}{2} \\ 0 & 0 & 1 & 0 & 0 \\ 0 & 0 & -1 & 0 & 0 \end{pmatrix}.$$

Writing out the equations for the conserved moment $n_0$ of Eqs. (B7)-(B9), the following equations can be obtained:

$$\epsilon^0:\ n_0^{(0)} = n_0^{eq}, \quad \text{(B10)}$$

$$\epsilon^1:\ \partial_{t_1} n_0^{(0)} + \partial_{x1} n_1^{(0)} + \partial_{y1} n_2^{(0)} = -\frac{\zeta_0}{\delta_t} n_0^{(1)}, \quad \text{(B11)}$$

$$\epsilon^2:\ \partial_{t_2} n_0^{(0)} + \partial_{t_1}\left[\left(1-\frac{\zeta_0}{2}\right)n_0^{(1)}\right] + \nabla_1 \cdot \left[\begin{pmatrix} 1-\zeta_1/2 & 0 \\ 0 & 1-\zeta_2/2 \end{pmatrix}\begin{pmatrix} n_1^{(1)} \\ n_2^{(1)} \end{pmatrix}\right] = -\frac{\zeta_0}{\delta_t} n_0^{(2)}, \quad \text{(B12)}$$

According to Eq. (B10), we have $n_0^{(k)} = 0$ ($k > 0$), then Eqs. (B11) and (B12) can be rewritten as follows:

$$\epsilon^1: \ \partial_{t_1} T + \partial_{x1}(u_x T) + \partial_{y1}(u_y T) = 0, \tag{B13}$$

$$\epsilon^2: \ \partial_{t_2} T + \nabla_1 \cdot \left[ \begin{pmatrix} 1-\zeta_1/2 & 0 \\ 0 & 1-\zeta_2/2 \end{pmatrix} \begin{pmatrix} n_1^{(1)} \\ n_2^{(1)} \end{pmatrix} \right] = 0. \tag{B14}$$

From Eq. (B8), we can get

$$n_1^{(1)} = -\frac{\delta_t}{\zeta_1} \left[ \partial_{t_1}(u_x T) + \frac{\varpi}{2} \partial_{x1} T \right], \tag{B15}$$

$$n_2^{(1)} = -\frac{\delta_t}{\zeta_2} \left[ \partial_{t_1}(u_y T) + \frac{\varpi}{2} \partial_{y1} T \right]. \tag{B16}$$

Substituting Eqs. (B15) and (B16) into Eq. (B14), we have

$$\epsilon^2: \ \partial_{t_2} T = \nabla_1 \cdot \left[ \delta_t \begin{pmatrix} \frac{1}{\zeta_1}-0.5 & 0 \\ 0 & \frac{1}{\zeta_2}-0.5 \end{pmatrix} \begin{pmatrix} \partial_{t_1}(u_x T) + \frac{\varpi}{2} \partial_{x1} T \\ \partial_{t_1}(u_y T) + \frac{\varpi}{2} \partial_{y1} T \end{pmatrix} \right]. \tag{B17}$$

Combining Eq. (B13) with Eq. (B17), the following macroscopic equation can be obtained

$$\partial_t T + \nabla \cdot (\mathbf{u} T) = \nabla \cdot \left[ \alpha \nabla T + \delta_t (\zeta_\alpha^{-1} - 0.5) \epsilon \partial_{t_1} (\mathbf{u} T) \right], \tag{B18}$$

where $\alpha = c_{sT}^2 (\zeta_\alpha^{-1} - 0.5) \delta_t$ is the thermal diffusivity ($\zeta_1 = \zeta_2 = \zeta_\alpha$), in which $c_{sT}^2 = c^2 \varpi/2 = \varpi/2$ ($c_{sT}$ is the lattice sound speed of the D2Q5 model). As compared with the temperature equation (3), Eq. (B18) contains a deviation term $\nabla \cdot \left[ \delta_t (\zeta_\alpha^{-1} - 0.5) \epsilon \partial_{t_1} (\mathbf{u} T) \right]$. For incompressible thermal flows, this deviation term can be ignored, and then the temperature equation (3) can be recovered. However, in certain cases (e.g., the pseudopotential multiphase LB model is adopted to solve the flow field) this deviation term may lead to considerable numerical errors [9], and then such a deviation term must be eliminated in simulations. The approach in Ref. [10] can be used to eliminate the deviation term based on the D2Q5 lattice model.